\documentclass[preprint,12pt,onecolumn]{elsarticle}

\usepackage{amssymb}
\usepackage{lipsum}

\usepackage{amsmath}
\usepackage{hyperref}

\hypersetup{
    colorlinks=true
}

\journal{the Proceedings of XQCD 2025}

\begin{document}

\begin{frontmatter}



\title{First Demonstration that Quark-Gluon Plasma has a Nonzero Resolution Length}


\author[a]{Arjun Srinivasan Kudinoor \fnref{speaker}}
\author[b,c,d]{Daniel Pablos}
\author[a]{Krishna Rajagopal}

\fntext[speaker]{Speaker at XQCD 2025.}

\affiliation[a]{
    organization={Center for Theoretical Physics -- a Leinweber Institute, Massachusetts Institute of Technology},
    city={Cambridge},
    state={MA},
    postcode={02139},
    country={USA}
}


\affiliation[b]{
    organization={IGFAE, Universidade de Santiago de Compostela},
    addressline={E-15782 Galicia-Spain}
}

\affiliation[c]{
    organization={Departamento de Fisica, Universidad de Oviedo},
    addressline={Avda. Federico Garcia Lorca 18},
    city={33007 Oviedo},
    country={Spain}
}

\affiliation[d]{
    organization={Instituto Universitario de Ciencias y Tecnologias Espaciales de Asturias (ICTEA)},
    addressline={Calle de la Independencia 13},
    city={33004 Oviedo},
    country={Spain}
}

%


\begin{abstract}
We report on our investigation~\cite{Kudinoor:2025gao} of how recent jet substructure measurements constrain the resolution length $L_{\rm res}$ of the quark-gluon plasma formed in heavy-ion collisions. $L_{\rm res}$ is defined such that high-energy partons within a jet shower are resolved by the medium if and only if they are separated by a distance greater than $L_{\rm res}$. Using the Hybrid Model, we reproduce ALICE data on the scaled Soft Drop angle $\theta_g$ for $R=0.2$ charged-particle jets and ATLAS data on the Hard Group angle $\Delta R_{12}$ for $R=1$ jets reclustered from skinny $R = 0.2$ inclusive subjets.
We find that the narrowing of the $\theta_g$-distribution in PbPb collisions observed by ALICE and the suppression of $R=1$ jets with multiple skinny subjets in PbPb collisions observed by ATLAS rule out $L_{\rm res} = \infty$, where each entire parton shower loses energy to the plasma coherently as if it were a single colored object. We then compare Hybrid Model calculations to ATLAS measurements of $R_{\rm AA}$ of $R=1$ jets reclustered from $R=0.2$ subjets, as a function of the Soft Drop angle $dR_{12}$ obtained by grooming all charged-particle tracks associated with each $R=1$ jet. We demonstrate, for the first time, that the ATLAS data is inconsistent with $L_{\rm res} = 0$, where the plasma resolves every splitting in a parton shower. Our results agree best with the data when QGP possesses a finite, nonzero $L_{\rm res}\sim (1-2)/(\pi T)$.
\end{abstract}




\end{frontmatter}




\section{The Hybrid Strong/Weak Coupling Model of Jet Quenching}
\label{introduction}

The high-temperature, strongly coupled phase of QCD matter called quark-gluon plasma (QGP) is formed in relativistic heavy-ion collisions. A particularly compelling way to access information about the microscopic
structure and dynamics of QGP is via the study of heavy-ion collision events in which high-energy jets are produced and quenched. We focus on studying how substructure-dependent suppression of jets in PbPb collisions can teach us about the resolution length $L_{\rm res}$ of QGP, which is defined such that if two partons in a jet are separated by a distance less than $L_{\rm res}$, they interact coherently with the plasma as if they were a single colored object; if and only if the two partons are separated by a distance greater than $L_{\rm res}$ will they interact independently with the plasma.

The hybrid strong/weak coupling model, or simply the Hybrid Model, is a theoretical framework for jet quenching in heavy-ion collisions.
Jet showers emerge from parton splittings that are determined using the high-$Q^2$ perturbative DGLAP evolution equations, implemented using PYTHIA8. As partons in a jet shower propagate through the drop of QGP, they undergo nonperturbative, soft momentum-exchanges with the medium, which cause these partons to lose energy. In the Hybrid Model, each parton in a jet shower loses energy to the plasma as determined by a holographic energy-loss formula,
detailed in Refs.~\cite{Chesler:2014jva,Chesler:2015nqz,Casalderrey-Solana:2014bpa,Casalderrey-Solana:2016jvj,Hulcher:2017cpt,Casalderrey-Solana:2018wrw,
Casalderrey-Solana:2019ubu,
Bossi:2024qho,Kudinoor:2025ilx,Kudinoor:2025gao}. The energy-loss formula includes one dimensionless parameter, denoted $\kappa_{\rm sc}$, that governs the strength of the
interaction between the jet parton and the QGP. $\kappa_{\rm sc}$ is obtained by fitting to data, as described in Refs.~\cite{Casalderrey-Solana:2018wrw,Kudinoor:2025gao}. Since energy and momentum must be conserved, the momentum and energy lost by each parton is deposited into the plasma, exciting a hydrodynamic wake within the droplet of QGP. The prescription for generating these jet-induced wakes and its implementation is described in Refs.~\cite{Casalderrey-Solana:2016jvj,Bossi:2024qho,Kudinoor:2025ilx}.


In the Hybrid Model, the QGP resolution length parameter $L_{\rm res}$ is implemented as detailed in Refs.~\cite{Hulcher:2017cpt,Casalderrey-Solana:2019ubu,Kudinoor:2025ilx,Kudinoor:2025gao}.
Two partons within the same parton shower lose energy independently to the medium, and excite a wake in the medium independently, if and only if they are separated by a distance larger than $L_{\rm res}$.
In these Proceedings, we review the main results of 
Ref.~\cite{Kudinoor:2025gao}, where we compared experimental data to Hybrid Model calculations with four values of $L_{\rm res}$: 0, $1/(\pi T)$, $2/(\pi T)$, and $\infty$, where $T$ is the temperature of the plasma. $L_{\rm res} = 0$ corresponds to fully incoherent energy-loss, where every parton in each jet shower is resolved. $L_{\rm res} = \infty$ corresponds to fully coherent energy-loss, where each entire jet shower loses energy and excites a wake as if it were a single color-charged parton. The physical motivation for choosing $1/(\pi T)$ and $2/(\pi T)$ for finite, nonzero values of $L_{\rm res}$ is that a QGP resolution length for massless, light-like partons in jets is akin to the Debye screening length $\lambda_D$ for infinitely massive, static color-charges within QGP --- and both weakly coupled perturbative calculations applied at $\alpha_s \sim 1/3$ and strongly coupled holographic calculations~\cite{Bak:2007fk} yield values of $\lambda_D$ around $1/(\pi T)$~\cite{Kudinoor:2025gao}. The connection between $L_{\rm res}$ and $\lambda_{D}$ is described more precisely in Refs.~\cite{Hulcher:2017cpt,Casalderrey-Solana:2019ubu,Kudinoor:2025ilx, Kudinoor:2025gao}.

\section{Constraining $L_{\rm res}$ From Above}

\subsection{Constraining $L_{\rm res}$ using Soft Drop}

Among various procedures to study jet substructure, the Soft Drop grooming algorithm ~\cite{Larkoski:2014wba}  (aka Soft Drop) is especially well-suited for probing the QGP resolution length. It identifies the first hard branching within an anti-$k_t$ jet of radius $R$ by first reclustering the jet's constituents using the Cambridge–Aachen (CA) algorithm. Stepping through this reclustering sequence, the Soft Drop algorithm selects the first splitting that satisfies
\begin{equation}
    z > z_{\rm cut} \left(\frac{R_{12}}{R}\right)^{\beta},
\end{equation}
where $z_{\rm cut}$ and $\beta$ are tunable parameters, $z$ is the fraction of $p_T$ carried by the subleading prong in the splitting, and $R_{12}$ is the angle between the leading and subleading prongs. The first hard splitting that satisfies this condition defines the Soft Drop angle $R_g \equiv R_{12}$ and the scaled Soft Drop angle $\theta_g \equiv R_g/R$. Since the Soft Drop grooming algorithm identifies the first hard splitting in a jet and quantifies its opening angle, it offers a clear probe of the medium’s ability to resolve nearby high-energy prongs within a jet.

\begin{figure}[t]
\begin{center}
\includegraphics[scale=0.9]{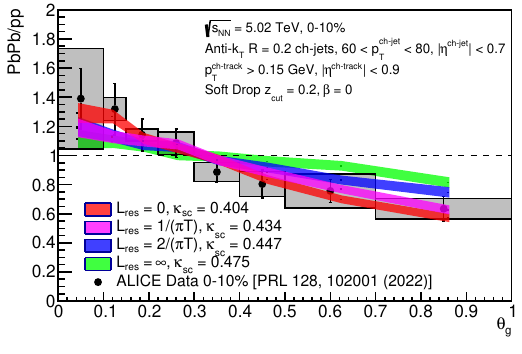}
\caption{
    \label{fig:softdrop}Ratio of the differential cross sections of jets in PbPb collisions to jets in pp collisions, as a function of the scaled Soft Drop angle $\theta_g$ for anti-$k_t$ $R = 0.2$ charged-particle jets with $60 < p_T^{\rm ch-jet} < 80$ GeV and $|\eta^{\rm ch-jet}| < 0.7$, reconstructed using the Soft Drop grooming procedure with parameters $z_{\rm cut} = 0.2$ and $\beta = 0$. The colored bands show the results of Hybrid Model calculations with $L_{\rm res} = 0$ (red), $1/(\pi T)$ (purple), $2/(\pi T)$ (blue) and $\infty$ (green). ALICE experimental measurements from Ref.~\cite{ALargeIonColliderExperiment:2021mqf} are depicted using point markers, upon which the vertical bars indicate statistical uncertainties and the shaded boxes indicate systematic uncertainties extracted from Ref.~\cite{ALargeIonColliderExperiment:2021mqf}.
}
\end{center}
\end{figure}

In Ref.~\cite{Kudinoor:2025gao}, we analyzed the differential cross sections of jets in PbPb collisions as a function of the scaled Soft Drop angle $\theta_g$, for different values of $L_{\rm res}$. Fig.~\ref{fig:softdrop} shows the ratio of these differential cross sections of jets in PbPb collisions to jets in pp collisions, as a function of $\theta_g$ for anti-$k_t$ $R = 0.2$ charged-particle jets with $60 < p_{T}^{\rm ch-jet} < 80$ GeV and $|\eta^{\rm ch-jet}| < 0.7$, reconstructed using the Soft Drop grooming algorithm with $z_{\rm cut} = 0.2$ and $\beta = 0$. We compare our Hybrid Model calculations to ALICE experimental measurements from Ref.~\cite{ALargeIonColliderExperiment:2021mqf}. We note that regardless of $L_{\rm res}$, both our model calculations and the experimental data show that jets within this range of $p_T$ in PbPb collisions have, on average, smaller $\theta_g$ than jets with the same reconstructed $p_T$ in pp collisions. There are multiple reasons for this, as enumerated below.

Since the production cross section for jets is a steeply falling function of $p_T$, the sample of jets within a reconstructed-$p_T$ bin are biased towards those that lost less energy. This ``survivor bias" in PbPb collisions reduces the fraction of jets with larger values of $\theta_g$ in three ways.
(i) Gluon-initiated jets tend to lose more energy than quark-initiated jets, 
regardless of whether QGP can resolve differences in the substructure of their jet showers. And, high-energy gluons tend to split at wider angles than high-energy quarks. So, a sample of jets with a given $p_T$ after quenching will have a smaller fraction of gluon-initiated jets than in vacuum, meaning a smaller $\theta_g$ on average, regardless of the value of $L_{\rm res}$.
(ii) The progressively steeper narrowing of the Soft Drop angle observed in the Hybrid Model curves with smaller $L_{\rm res}$ in Fig.~\ref{fig:softdrop} arises because, within the Hybrid Model, jets that contain multiple resolved sources of energy-loss lose more energy than jets with the same initial $p_T$ but with fewer resolved sources of energy-loss. The smaller the value of $L_{\rm res}$ the sooner that subjets separated by an angle $R_g = R \theta_g$ will be resolved. So, for any finite value of $L_{\rm res}$, jets lose more energy if their $\theta_g$ is larger, and jets with a given $\theta_g$ lose more energy if $L_{\rm res}$ is smaller. Since jets with smaller $\theta_g$ tend to lose less energy, the jet survivor bias in PbPb collisions reduces the fraction of jets with larger values of $\theta_g$ when $L_{\rm res}$ is finite and makes the shrinking of $\theta_g$ progressively steeper at smaller values of $L_{\rm res}$ as seen in Fig.~\ref{fig:softdrop}.
(iii) As described in Ref.~\cite{Casalderrey-Solana:2019ubu}, jets with larger $\theta_g$ have, on average, a greater number of subjets, i.e. a greater number of resolved sources of energy-loss when $L_{\rm res}$ is finite. So, those jets that survive the jet-$p_T$ cut will be biased towards those with fewer subjets that are resolved by the medium. The number of such resolved subjets increases with decreasing $L_{\rm res}$. This also contributes to the progressively steeper shrinking of $\theta_g$ at smaller values of $L_{\rm res}$.

Furthermore, since jets in PbPb collisions lose energy to the plasma, imposing the same reconstructed–jet-$p_{T}$ selection in PbPb and in pp events preferentially selects higher $p_{T}^{\rm init}$ progenitor partons of jets in the heavy–ion data set. In the collinear approximation used in the Soft Drop grooming algorithm~\cite{Larkoski:2014wba}, the angle of a 1-to-2 splitting is
$R_{12}=k_T/\left(z \, p_T^{\rm init}\right)$, where $k_T$ is the relative transverse momentum scale of the splitting. This means that at fixed $(k_{T}, z)$, higher-$p_T$ progenitors of jets result in splittings with narrower $R_{12}$, and therefore smaller $\theta_g$, regardless of the ability of the medium to resolve these splittings.

In Fig. 1 we observe that the narrowing of the $\theta_g$ distribution in PbPb collisions relative to that in pp collisions, as seen in the ALICE data, is inconsistent with our Hybrid Model calculations for $L_{\rm res}=\infty$, where no substructure within a jet is resolved by the medium. The comparison between the ALICE experimental data and our calculations therefore suggests that QGP does in fact resolve jet substructure. However, given the current experimental uncertainties, the data do not allow us to distinguish between the three finite values of $L_{\rm res}$ that we have investigated. The sensitivity to $L_{\rm res}$ appears to increase for jets with more widely separated prongs (larger $\theta_g$), suggesting a natural next step: to explore methods for studying jets whose prongs are separated by even larger angles.

\subsection{Constraining $L_{\rm res}$ using Hard Group}

\begin{figure} [t]
    \begin{center}
    \includegraphics[scale=0.45]{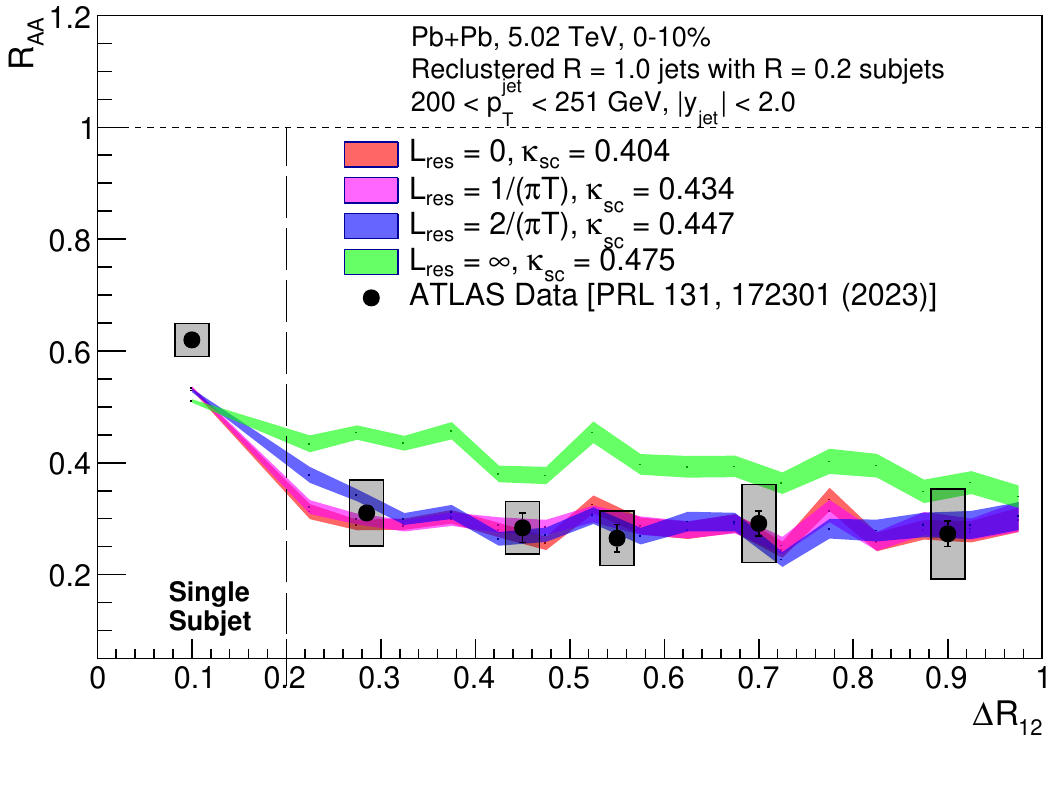}
    \caption{$R_{\rm AA}$ as a function of $\Delta R_{12}$ for reclustered large-radius $R = 1.0$ jets with $200 < p_T < 251$ GeV. The left-most bin denotes the value of $R_{\rm AA}$ for $R = 1.0$ jets that contain only a single skinny subjet, placed arbitrarily at 0.1; all bins with $\Delta R_{12}\geq 0.2$ show $R_{\rm AA}$ for $R=1.0$ jets composed of multiple skinny subjets with the hardest splitting given by $\Delta R_{12}$, as described in the text. The colored bands show the results of Hybrid Model calculations with $L_{\rm res} = 0$ (red), $1/(\pi T)$ (purple), $2/(\pi T)$ (blue) and $\infty$ (green).
    ATLAS experimental measurements from Ref.~\cite{ATLAS:2023hso} are depicted using point markers. The vertical bars on ATLAS' experimental data points indicate statistical uncertainties and the shaded boxes indicate systematic 
    uncertainties~\cite{ATLAS:2023hso}.}
    \label{fig:deltaR}
    \end{center}
\end{figure}

In Ref.~\cite{ATLAS:2023hso}, the ATLAS Collaboration introduced a procedure of reconstructing jets, which we refer to as the \textit{Hard Group} prodcedure, that enables access to large angular separations between prongs within large-radius jets. They first reconstructed anti-$k_t$ $R = 0.2$ jets with $|\eta| < 3.0$ and $p_T > 35$ GeV. They then used these $R = 0.2$ jets, which we refer to as ``skinny subjets", as the constituents for reconstructing anti-$k_t$ $R = 1$ jets with $|y| < 2.0$ and $p_T > 158$~GeV. Schematically, hard skinny subjets are grouped together into large-radius $R = 1$ jets (hence the name Hard Group). Finally, the $R = 1$ jets were reclustered using the $k_t$-recombination algorithm to obtain the Hard Group angle
\begin{equation}
    \Delta R_{12} \equiv \sqrt{\Delta y_{12}^2 + \Delta \phi_{12}^2},
\end{equation}
defined as the angular separation between the two subjet-constituents involved in the final reclustering step of the $R = 1$ jet. Since the $k_t$ algorithm tends to combine the hardest constituents of a jet last, $\Delta R_{12}$ corresponds to the angle of the hardest splitting within a large-radius $R = 1.0$ jet. Since each large-radius $R=1$ jet is composed from $R=0.2$ skinny subjets, the Hard Group angle $\Delta R_{12}$ has to be greater than 0.2. Thus, the Hard Group procedure of reconstructing $R=1$ jets from $R = 0.2$ subjets probes splittings with angular separations wider than the largest possible Soft Drop angle of an $R = 0.2$ jet.

In Ref.~\cite{Kudinoor:2025ilx} we analyzed the suppression of large-radius $R = 1$ jets constructed using the Hard Group algorithm as a function of their transverse momentum and $\Delta R_{12}$, for different values of $L_{\rm res}$. Consistent with our findings in the previous Section, we found that the ATLAS experimental data rules out $L_{\rm res} = \infty$, where an entire parton shower loses energy coherently to the medium as a single colored object. This is clearly seen in Fig.~\ref{fig:deltaR}, which shows $R_{\rm AA}$ as a function of $\Delta R_{12}$ --- namely the reduction in the number of jets in a given bin of $dR_{12}$ in PbPb collisions relative to that in the corresponding number of pp collision --- for large-radius $R = 1$ jets with $200 < p_T < 251$ GeV, reconstructed using the Hard Group procedure described above.

We observe that for finite $L_{\rm res}$ and in the ATLAS data, large-radius jets with multiple subjets ($\Delta R_{12} > 0.2$) are more suppressed than large-radius jets with a single subjet. This is because jets that contain multiple resolved sources of energy-loss lose more energy than jets with the same initial $p_T$ but with a single source of energy-loss. In the case of infinite $L_{\rm res}$, no splittings are ever resolved by the medium, suggesting that a large-radius jet with multiple subjets should be just as suppressed as a large-radius jet with a single subjet. However, Fig.~\ref{fig:deltaR} shows that large-radius jets with multiple subjets are more suppressed than large-radius jets with a single subjet, even when $L_{\rm res} = \infty$. We understand this to be the result of initial-state radiation, as is explained in detail in Ref.~\cite{Kudinoor:2025ilx}.

We note, from Fig.~\ref{fig:deltaR}, that the ATLAS experimental data is inconsistent with the Hybrid Model calculation with $L_{\rm res}=\infty$, indicating that $L_{\rm res}$ is finite. This is consistent with what we found in the previous section --- namely that the ALICE measurements in Fig.~\ref{fig:softdrop} rule out an infinite QGP resolution length. Furthermore, all Hybrid Model calculations with
finite resolution length in Fig.~\ref{fig:deltaR} describe
the ATLAS Hard Group measurements across the range $0.2<\Delta R_{12}<1.0$. This suggests~\cite{Kudinoor:2025ilx, Kudinoor:2025hjc} that, in order to constrain the value of $L_{\rm res}$ further, the most interesting range of angular separation between hard prongs within jets could be around 0.2. This elusive angular region is larger than what is accessible in the ALICE Soft Drop measurements in Fig.~\ref{fig:softdrop} and smaller than what is accessible in the ATLAS Hard Group measurements in Fig.~\ref{fig:deltaR}.

\section{Constraining $L_{\rm res}$ From Below}

\begin{figure*}[t]
\begin{center}
\includegraphics[scale=0.75]{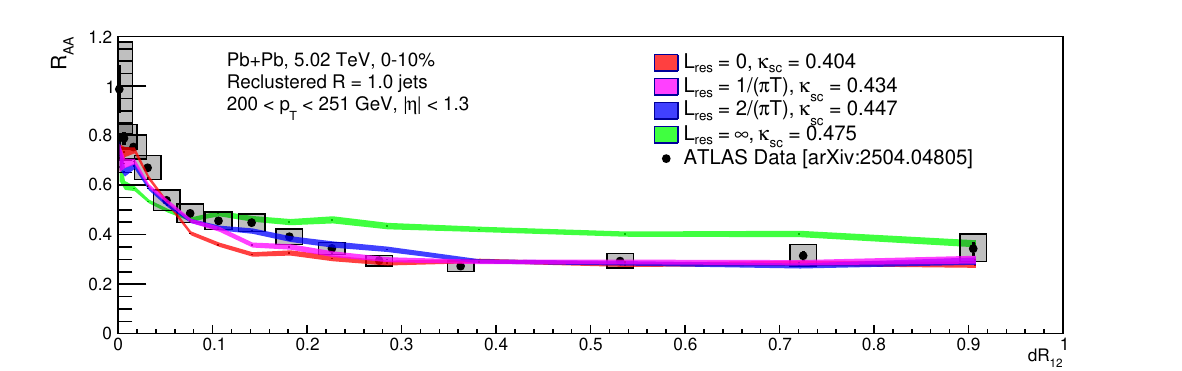}
\caption{
    \label{fig:dr12}$R_{\rm AA}$ as a function of $dR_{12}$ for $R = 1.0$ jets with $200 < p_T < 251$ GeV and $|\eta| < 1.3$ constructed from skinny $R = 0.2$ subjets using the Hard Group procedure, whose associated charged-particle tracks were then reclustered using the $k_t$ algorithm and the Soft Drop grooming procedure with parameters $z_{\rm cut} = 0.15$ and $\beta = 0$. The colored bands show the results of Hybrid Model calculations with $L_{\rm res} = 0$ (red), $1/(\pi T)$ (purple), $2/(\pi T)$ (blue) and $\infty$ (green). ATLAS experimental measurements from Ref.~\cite{ATLAS:2025lfb} are depicted using point markers, upon which the vertical bars indicate statistical uncertainties and the shaded boxes indicate systematic uncertainties extracted from Ref.~\cite{ATLAS:2025lfb}.
}
\end{center}
\end{figure*}

The Soft Drop grooming algorithm employed by ALICE on charged-particle tracks provided fine granularity in measuring the angular separation of a hard splitting within a jet, but was restricted by the small jet radius. Conversely, the Hard Group procedure used by ATLAS with calorimetric jets accessed large-angle splittings in $R=1$ jets, though its resolution was limited by the size of the skinny subjets. To overcome both limitations, ATLAS introduced~\cite{ATLAS:2025lfb} a new observable, $dR_{12}$, which interpolates between the Soft Drop $R_g$ at small values and the Hard Group $\Delta R_{12}$ at large values.

Following ATLAS, $dR_{12}$ is constructed as follows. $R=1$ jets with $p_T > 158$~GeV and $|\eta| < 1.3$ are first reconstructed from skinny $R=0.2$ subjets with $p_T > 35$ GeV and $|\eta| < 3.0$ via the Hard Group procedure. Instead of reclustering only the subjets, all charged-particle tracks with $p_T > 4$ GeV associated with the large-radius jet are identified using a ghost-association procedure (described in detail in both Refs.~\cite{ATLAS:2025lfb} and~\cite{Kudinoor:2025gao}) and then reclustered using the $k_t$ algorithm with $R = 2.5$. The Soft Drop grooming condition is applied
with $z_{\rm cut}=0.15$ and $\beta=0$ to the reclustering history, which yields a Soft Drop angle $dR_{12} \equiv R_g$ for each large-radius $R = 1$ jet.

In Ref.~\cite{Kudinoor:2025gao} we analyzed the suppression of such $R = 1$ jets as a function of $dR_{12}$, for different values of $L_{\rm res}$. Figure~\ref{fig:dr12} shows ATLAS measurements of $R_{\rm AA}$ as a function of $dR_{12}$ for $R=1$ jets with $200 < p_T < 251$~GeV, compared with Hybrid Model calculations at several values of $L_{\rm res}$. This observable extends the Soft Drop (Hard Group) analysis to large (small) angles while retaining sensitivity at small (large) angles. Consistent with the Soft Drop and Hard Group analyses presented in these Proceedings, the ATLAS data for large-radius jet suppression as a function of $dR_{12}$ rule out a picture of fully coherent energy-loss within jets due to an infinite resolution length. Strikingly, for the first time as seen in experimental data, the data points in Fig.~\ref{fig:dr12} also disfavor a picture of fully incoherent energy-loss within jets due to a zero-valued resolution length! Instead, the ATLAS data points are best described by finite nonzero values of $L_{\rm res}$, with $L_{\rm res}=2/(\pi T)$ providing a somewhat better description than $1/(\pi T)$. This provides a clear indication that QGP has a finite, nonzero resolution length.  And, this demonstration establishes a clear path toward drawing a definitive conclusion with higher statistics data to come.

\subsection{A Next Step: Elastic Scatterings}

\begin{figure*}[t]
\begin{center}
\includegraphics[scale=0.75]{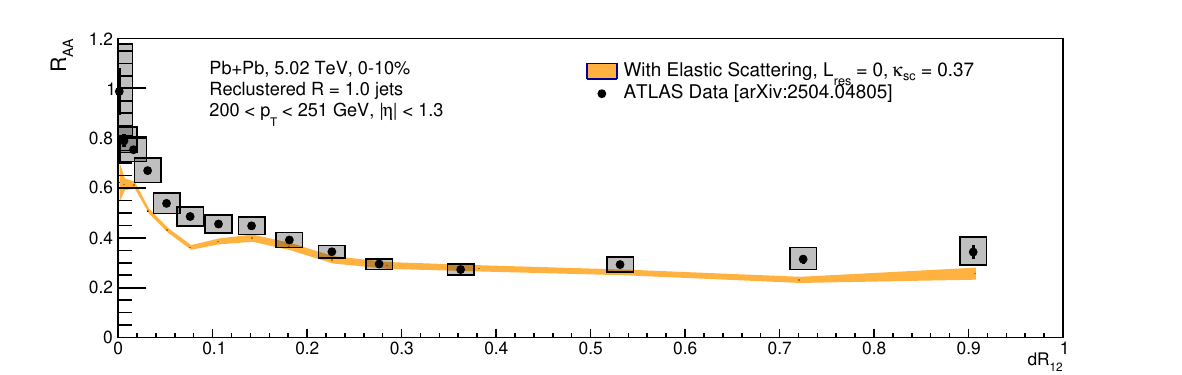}
\caption{
    \label{fig:dr12-elastic}$R_{\rm AA}$ as a function of $dR_{12}$ for $R = 1.0$ jets with $200 < p_T < 251$ GeV and $|\eta| < 1.3$ constructed from skinny $R = 0.2$ subjets using the Hard Group procedure, whose associated charged-particle tracks were then reclustered using the $k_t$ algorithm and the Soft Drop grooming procedure with parameters $z_{\rm cut} = 0.15$ and $\beta = 0$. The orange band shows the results of Hybrid Model calculations with elastic $2 \rightarrow 2$ scatterings, and $L_{\rm res} = 0$. ATLAS experimental measurements from Ref.~\cite{ATLAS:2025lfb} are depicted using point markers, upon which the vertical bars indicate statistical uncertainties and the shaded boxes indicate systematic uncertainties extracted from Ref.~\cite{ATLAS:2025lfb}.
}
\end{center}
\end{figure*}

Up to this point, our discussion of $R_{\rm AA}$ as a function of $dR_{12}$ has focused exclusively on the role of the resolution length $L_{\rm res}$, without considering the effects of other phenomena like elastic $2 \rightarrow 2$ scatterings between high-energy partons within jets and quark- and gluon-like quaspiarticles in the QGP medium. Within the Hybrid Model, a jet-parton that
scatters is deflected, kicking a medium-parton, which recoils. As both of these partons propagate further through the medium they lose energy and momentum to the medium, exciting
wakes, as described in Ref.~\cite{Hulcher:2022kmn}. These elastic scatterings, which can deflect and modify the angular distribution of high-energy partons within jets, have been investigated in Refs.~\cite{Hulcher:2022kmn, Pablos:2024muu, Bossi:2025kac} but have not yet been incorporated into the Hybrid Model with nonzero $L_{\rm res}$.

Figure~\ref{fig:dr12-elastic} shows that although the $dR_{12}$ observable is sensitive to the effects of elastic scatterings, our Hybrid Model calculations with $L_{\rm res} = 0$ still fail to describe the ATLAS data in the region $dR_{12}\sim0.05{-}0.2$, even after we include elastic scatterings. In particular, our Hybrid Model calculations overpredict the suppression of jets with narrow substructure. This reinforces the conclusion that a fully incoherent picture of energy-loss is disfavored by the experimental data, while it also provides strong motivation to extend the Hybrid Model to include elastic scatterings in the case of finite, nonzero $L_{\rm res}$. Incorporating these effects will be a crucial next step, as many jet substructure observables are expected to be particularly sensitive to both elastic scatterings and the QGP resolution length.

\section*{Acknowledgements}
Research supported in part by the U.S.~Department of Energy, Office of Science, Office of Nuclear Physics under grant Contract Number DE-SC0011090, by the European Union's Horizon 2020 research and innovation program under the Marie Sk\l odowska-Curie grant agreement No 101155036 (AntScat), by the European Research Council project ERC-2018-ADG-835105 YoctoLHC, by the Spanish Research State Agency under project 
PID2020-119632GB-I00, by Xunta de Galicia (CIGUS Network of Research Centres) and the European Union, and by Unidad de Excelencia Mar\'ia de Maetzu under project CEX2023-001318-M.
ASK is supported by the National Science
Foundation Graduate Research Fellowship Program under Grant No.~2141064. DP is supported by the Ram\'on y
Cajal fellowship RYC2023-044989-I  funded by Spain’s Ministry of Science, Innovation and
Universities.
KR acknowledges the hospitality of the CERN Theory Department and the Aspen Center for Physics, which is supported by National Science Foundation grant PHY-2210452.

\bibliographystyle{elsarticle-num} 
\bibliography{bibliography}






\end{document}